\documentclass{article}
\usepackage{spconf,amsmath,graphicx,hyperref}
\usepackage{cite}
\usepackage{amsmath,amssymb,amsfonts}
\usepackage{algorithmic}
\usepackage{graphicx}
\usepackage{textcomp}
\usepackage{xcolor}
\usepackage{booktabs}
\usepackage{dcolumn}
\usepackage{url}
\usepackage{enumitem}
\usepackage{multirow}
\usepackage{arydshln}

\title{Leveraging Whisper Embeddings for Audio-based Lyrics Matching}
\name{
  Eleonora Mancini$^1$ \qquad
  Joan Serrà$^2$ \qquad
  Paolo Torroni$^1$ \qquad
  Yuki Mitsufuji$^{2,3}$
}

\address{
  $^1$ DISI, University of Bologna \quad
  $^2$ Sony AI \quad
  $^3$ Sony Group Corporation
}

\begin{document}
\ninept
\maketitle

\begin{abstract}
Audio-based lyrics matching can be an appealing alternative to other content-based retrieval approaches, but existing methods often suffer from limited reproducibility and inconsistent baselines. In this work, we introduce WEALY, a fully reproducible pipeline that leverages Whisper decoder embeddings for lyrics matching tasks. WEALY establishes robust and transparent baselines, while also exploring multimodal extensions that integrate textual and acoustic features. Through extensive experiments on standard datasets, we demonstrate that WEALY achieves a performance comparable to state-of-the-art methods that lack reproducibility. In addition, we provide ablation studies and analyses on language robustness, loss functions, and embedding strategies. This work contributes a reliable benchmark for future research, and underscores the potential of speech technologies for music information retrieval tasks.
\end{abstract}


\section{Introduction}

Audio-based lyrics matching represents a challenge in music information retrieval, with potential applications spanning copyright protection, music discovery, and creative assistance. For copyright enforcement, identifying songs with similar lyrical content is essential as audio-related aspects and lyrical composition aspects are typically protected under different legal frameworks, the latter requiring specialised tools for detecting textual similarities and potential infringement. In music discovery systems, lyrics-based matching enables users to find thematically related songs, or songs that share semantic connections beyond acoustic similarities. For creative professionals, lyrics matching serves as an inspiration tool that can identify existing works with similar themes or lyrical structures, enabling songwriters to explore new creative directions while ensuring they avoid unintentional plagiarism. 

A straightforward approach to lyrics matching may rely on pre-existing text or transcriptions. However, text is often unavailable, incomplete, or restricted by copyright. A more general challenge for lyrics matching approaches is to capture multilingual semantic relationships, where the same content expressed in different languages should be recognised as similar, and to account for semantic nuances across linguistic and contextual variations. Furthermore, existing methods either employ outdated techniques, lack reproducibility, or are evaluated on limited datasets that fail to establish robust benchmarks for the field (see below). These issues are compounded by copyright restrictions that limit access to lyrical databases, inherent limitations of traditional automatic speech recognition systems (ASR) when processing musical content, scalability requirements for real-world deployment across large music catalogs, and the impracticality of relying on manual annotation platforms like Genius, which cannot scale to cover the vast expanse of global music production. The evaluation of lyrics matching systems is as well hindered by data availability and scarcity issues. 

In this work, we introduce WEALY (\textbf{W}hisper \textbf{E}mbeddings for \textbf{A}udio-based \textbf{LY}rics matching), a fully reproducible end-to-end pipeline that leverages Whisper~\cite{whisper} decoder embeddings, to address the abovementioned limitations and challenges, eliminating the dependency on text data or pre-existing transcriptions while providing robust and transparent baselines. This choice is further supported by recent findings which show that Whisper-based transcriptions can already provide strong results compared to annotated ground truth~\cite{Balluff2024LyricCovers}, reinforcing the value of end-to-end audio-based approaches. To evaluate our approach, we employ musical version identification (MVI)~\cite{yesiler_audio-based_2021} as a proxy task, motivated by the challenges of data scarcity in direct lyrics matching evaluation. With it, we establish comprehensive benchmarks across three different datasets and achieve a competitive performance compared to strong baselines. We additionally conduct extensive ablation studies to analyze the impact of different losses and pooling strategies. Our ongoing work also explores multimodal extensions that integrate textual and acoustic features. We release both the code and the models' checkpoints\footnote{\scriptsize{\url{https://github.com/helemanc/audio-based-lyrics-matching}}} to support transparency and reproducibility.  


\section{Related Work}

The specific problem of lyrics matching remains largely unexplored. Prior text-based approaches~\cite{knees2005multiple, patra2017retrieving} face critical limitations, including transcription errors, spelling variations and structural inconsistencies. 
These approaches require extensive manual data collection and preprocessing, illustrating their limited feasibility and motivating our end-to-end audio-based method that extracts semantic lyric representations directly from audio. 
The lack of large-scale, high-quality lyrics datasets justifies using MVI~\cite{yesiler_audio-based_2021} as a proxy task for large-scale evaluation, enabling the derivation of similarity labels without relying on curated texts.

While audio-based MVI systems~\cite{serra2025supervised, bytecover2, bytecover3} achieve strong retrieval performance using audio embeddings, lyrics-based pipelines remain limited. Early lyrics-based pipelines~\cite{vaglio-hal-03356164, correya2018large} combined singing voice recognition with string similarity measures or integrated metadata with tf-idf lyrics from Musixmatch, but were constrained by transcription errors and limited lyric coverage. Recent works~\cite{Balluff2024LyricCovers, DBLP:conf/ismir/DuLZLWLZ24} leverage advances in automatic speech recognition (ASR), particularly Whisper, and combine source separation, transcriptions, and various embeddings trained with multiple loss objectives. However, these approaches rely on highly complex, resource-heavy pipelines with limited transparency and reproducibility.

\begin{figure*}[t]
    \centering
    \includegraphics[width=\textwidth]{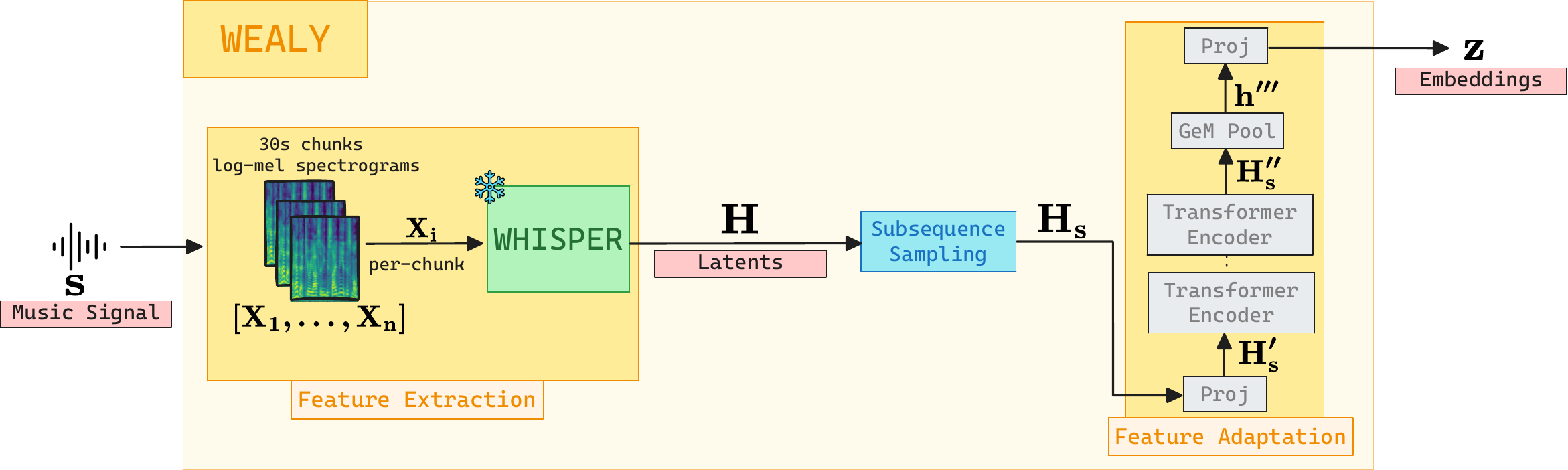}
    \caption{Overview of the WEALY architecture. Stage 1 extracts lyrical latents with Whisper, while stage 2 learns contextualized representations with a transformer encoder and projects them into a compact embedding space.}
    \label{fig:whisper-embeddings-extraction}
\end{figure*}

Recent advances in large-scale ASR models have enabled more direct extraction of lyric information from audio. While Whisper has been successfully applied to automatic lyrics transcription~\cite{DBLP:conf/ismir/ZhuoYPMLZLDFLBC23} and MVI~\cite{DBLP:conf/ismir/DuLZLWLZ24}, the latter again lacks transparency and reproducibility. 
Unlike prior work, we evaluate on multiple datasets and establish multiple transparent and reproducible lyrics-oriented baselines.

\section{Method}

\subsection{Problem Formulation}
We formalise audio-based lyrics matching as the task of identifying semantic and structural similarities between songs based on their lyrical content, where songs are considered similar if they share thematic connections, semantic overlaps, or structural patterns in their lyrics. Given the absence of dedicated datasets for direct lyrics matching evaluation, we employ
MVI as a proxy task, operating under the assumption that musical versions sharing lyrical content should provide a lower bound on detectable semantic similarities. 
\vspace{-1mm}
\subsection{Pipeline}
WEALY employs a two-stage pipeline design for end-to-end lyrics matching, based on feature extraction and adaptation. For feature extraction, we process raw audio through our Whisper decoder embeddings extraction pipeline, producing dense tensor representations that capture the semantic content of song lyrics without requiring intermediate text transcription. For feature adaptation, we input lyrical representations into  a transformer-based architecture trained with contrastive learning on the MVI task.


\vspace{1mm}\noindent\textbf{Feature Extraction ---} 
Our approach begins with audio preprocessing, converting to mono, resampling files to 
16\,kHz, and truncating to a maximum of 5\,min to 
balance computational efficiency with content coverage. Unlike some prior work that applies vocal source separation before feature extraction, we operate directly on the raw mixture. Since some results suggest that even with a strong separator Whisper's transcription quality only improves modestly~\cite{Balluff2024LyricCovers}, we avoid source separation in our pipeline. 

Whisper employs 
a classic transformer architecture with stacked encoder and decoder 
blocks, where attention mechanisms propagate information between 
components as the system processes an input audio file 
$\textbf{s}$, splitting it into $n$ 30-second overlapping chunks so that 
$\textbf{s} = [\textbf{s}_1, \ldots, \textbf{s}_n]$. 
From each chunk, Whisper extracts log-mel spectrograms $\textbf{X}_i \in \mathbb{R}^{t \times 128}$, yielding a sequence of spectrograms $\textbf{X}=[\textbf{X}_1, \ldots, \textbf{X}_n]$, where $t$ depends on the frame length (Fig.~\ref{fig:whisper-embeddings-extraction}). 
For each log-mel spectrogram $\textbf{X}_i$, the encoder processes the 
raw audio and captures positional information for spoken elements while 
preserving temporal structure. The decoder then leverages this encoded 
information to predict tokens (words) in an autoregressive manner, 
using both audio representations and previously predicted tokens to 
build contextual understanding throughout the sequence.

From each $\textbf{X}_i$, we extract decoder representations 
$\textbf{H}_i = [\textbf{h}_i^{(j)},\dots \textbf{h}_i^{(m_i)}]$, where $m_i$ is the number of per-chunk hidden decoder states or latents. Since Whisper processes one chunk at a time, we concatenate the hidden representations after each autoregressive pass over one chunk to obtain the final representation. After 
autoregressive decoding, we concatenate all latents from all chunks into a single matrix
$\textbf{H} = [\textbf{h}_1^{(1)}, \ldots \textbf{h}_1^{(m_1)}, \textbf{h}_2^{(1)}, \ldots \textbf{h}_n^{(m_n)}]$. This results in $\textbf{H} \in \mathbb{R}^{m \times d_w}$, where $m = \sum_{i=1}^{n} m_i$ is the total number of latents across all chunks and $d_w = 1280$ 
is the latent dimension of Whisper-turbo's 4-layer decoder. 
The length $m$ varies because Whisper produces variable-length sequences of hidden states, skipping over silent regions and generating outputs only for active segments.
Thus, the final number of 
hidden states depends on the audio content characteristics rather than 
being determined solely by the track duration or a fixed segmentation 
strategy.

This autoregressive process enables each latent $\textbf{h}$ to incorporate prior context, thereby supporting coherent transcriptions that accurately reflect linguistic dependencies and semantic relationships. Context resets are triggered by temperature thresholds during 
uncertain generation, creating distinct segments in the hidden-state 
sequence. We extract representations from Whisper’s final decoder layer 
before token sampling, capturing the model’s refined semantic 
understanding of lyrical content: the compressed knowledge about the 
song’s lyrics before committing to specific word predictions. What we 
obtain can be cast as `lyrics-aware Whisper latents', which we use as 
representations of the music tracks.  
Our implementation selects hidden states corresponding to the 
highest-scoring beam search sequence through modifications to the 
original Whisper codebase. Additionally, based on the results obtained by \cite{DBLP:conf/ismir/ZhuoYPMLZLDFLBC23}, we investigate initialisation 
prompting effects by comparing standard tokens against task-specific 
prompts such as \emph{``lyrics''} in order to assess their influence on 
representation quality (Sec.~\ref{sec:results}).  

\vspace{1mm}\noindent\textbf{Feature Adaptation ---} The second stage of WEALY employs a transformer-based architecture to learn similarity representations from the Whisper-derived lyrical latents (Fig.~\ref{fig:whisper-embeddings-extraction}). Given the feature sequence $\mathbf{H}$ extracted in the first stage, we sample random subsequences $\mathbf{H}_s \in \mathbb{R}^{k \times d_{w}}$ of fixed length $k=1500$ to ensure computational efficiency and to expose the model to diverse temporal segments of each track. This subsequence length was chosen empirically: shorter windows (e.g.,~$k=500$ or 1000) failed to capture sufficient lyrical context for semantic alignment, while longer windows ($k=2000$) degraded performance.

Each subsequence $\mathbf{H}_s$ is first projected into the model dimension $d_h$ through a linear layer, yielding $\mathbf{H}'_s \in \mathbb{R}^{k \times d_h}$. The sequence is then processed by a stack of transformer encoder blocks, which produce contextualized representations $\mathbf{H}''_s \in \mathbb{R}^{k \times d_h}$ without altering the sequence length. To obtain a temporally-collapsed vector suitable for generic similarity computation, we apply Generalized Mean (GeM) pooling~\cite{radenovic2018fine} across the temporal axis, resulting in a single representation $\mathbf{h}''' \in \mathbb{R}^{d_h}$. Finally, a linear projection maps $\mathbf{h}'''$ into the target embedding space, yielding a compact semantic representation $\mathbf{z}\in\mathbb{R}^{d_e}$. 
Our final configuration comprises $N=4$ transformer encoder blocks. Each block uses a self-attention 
model dimension of $d_h=768$ with 12 attention heads 
and a feed-forward MLP dimension of 1024. A dropout $p=0.1$ is applied. Input subsequences are first linearly projected to $d_{\text{model}}$, processed by the $N$ blocks, GeM-pooled over time, and finally projected to an embedding of size $d_e=512$.


To learn our representation $\textbf{z}$, we adopt a contrastive learning paradigm based on the NT-Xent (Normalized Temperature-scaled Cross-Entropy) loss~\cite{pmlr-v119-chen20j}. Let $\text{sim}(\mathbf{u}, \mathbf{v}) = \frac{\mathbf{u}^\top \mathbf{v}}{\|\mathbf{u}\|\|\mathbf{v}\|}$ denote the cosine similarity between vectors $\mathbf{u}$ and $\mathbf{v}$. For a positive pair of songs $i$ and $j$, the loss is defined as
\begin{equation*}
\mathcal{L}_{i,j} = -\log \frac{\exp(\text{sim}(\mathbf{z}_i, \mathbf{z}_j)/\tau)}{\sum_{k=1}^{2N} \mathbf{1}_{[k \neq i]} \exp(\text{sim}(\mathbf{z}_i, \mathbf{z}_k)/\tau)},
\end{equation*}
where $\tau=0.1$ is a temperature parameter and $\mathbf{1}_{[k \neq i]}$ is an indicator function that excludes self-similarity. The final loss is computed symmetrically across all positive pairs $(i,j)$ and $(j,i)$ in the mini-batch. This encourages embeddings of versions of the same song to cluster together while pushing apart those from different songs.

\subsection{Ablation Studies}
The architecture in Fig.~\ref{fig:whisper-embeddings-extraction} represents the final design we converged upon after 
extensive experimentation with different pooling strategies, projection 
schemes, transformer configurations, and loss functions. In our results we report on ablation studies across key design choices: loss function, pooling strategy, and multimodal fusion. 

\vspace{1mm}\noindent\textbf{Loss function --- }We compare NT-Xent against the triplet loss~\cite{Schroff_2015_CVPR}, which uses margin-based ranking between positive and negative pairs, and the CLEWS loss~\cite{serra2025supervised}, originally designed for multi-sequence similarity learning. 

\vspace{1mm}\noindent\textbf{Pooling strategy ---} We evaluate simple average pooling against GeM Pooling, which applies learnable power-mean pooling to emphasize informative temporal regions. We also explore using the canonical CLS token representation instead of pooling across timesteps and directly averaging Whisper embeddings and passing them through an MLP before contrastive training (bypassing the transformer encoder). 

\vspace{1mm}\noindent\textbf{Language-constrained representations ---} 
To assess the role of multilingual cues in Whisper, we force the model to transcribe all songs into English tokens exclusively before generating latents. This setting allows us to test whether removing cross-lingual information degrades retrieval performance and to quantify the benefit of retaining multilingual features in the embeddings.  


\vspace{1mm}\noindent\textbf{Multimodal fusion ---}  We also conduct a small multimodal fusion experiment, where WEALY (lyrics-aware) and CLEWS (audio-content) are combined via late fusion at the distance level. For each song pair $(i,j)$, we compute distances $\delta_\text{CLEWS}(\textbf{s}^{(i)},\textbf{s}^{(j)})$ and $\delta_\text{WEALY}(\textbf{s}^{(i)},\textbf{s}^{(j)})$, and define the combined distance as $\delta(\textbf{s}^{(i)},\textbf{s}^{(j)}) = \delta_\text{CLEWS}(\textbf{s}^{(i)},\textbf{s}^{(j)}) + \alpha\cdot \delta_\text{WEALY}(\textbf{s}^{(i)},\textbf{s}^{(j)})$. We use the original bpwr formulation for $\delta_\text{CLEWS}$~\cite{serra2025supervised}, the cosine distance for $\delta_\text{WEALY}$, and we set $\alpha = 1.5$ based on validation experiments.

\section{Experimental Setup }

    
    



\subsection{Datasets}

We evaluate WEALY on three publicly-available datasets: DiscogsVI-YT (DVI)~\cite{araz2024discogs}, SHS100k-v2 (SHS)~\cite{yu2020learning}, and LyricCovers2.0 (LYC)~\cite{Balluff2024LyricCovers}. All audio is processed at 16 kHz mono with 5-minute length constraints.
SHS represents a well-established reference dataset, although it exhibits an important bias towards unrealistically large version group sizes~\cite{serra2025supervised} and YouTube link dependencies limit the collection of all entries (we were able to collect 82\% of it). 
DVI addresses these limitations by providing a more recent dataset that is five times larger and with more realistic version group size distributions, offering improved validity.
LYC encompasses 78,862 tracks (54,301~covers and 24,561 originals) with annotated lyrics via \url{genius.com} links. In its original form, the dataset contains duplicated entries within version groups, since tracks are paired with both Whisper-generated and annotated transcriptions where available. This design reflects one of the original authors’ goals, namely to evaluate transcription accuracy and to compare the impact of different transcription sources on MVI. In our case, since the focus is on evaluating audio-derived lyric representations rather than transcription quality, we reduced the dataset to unique clique–version pairs while retaining only the metadata needed for retrieval.
The dataset spans 80 languages (82.6\%~English) with 95\% of musical versions maintaining the original language, making it relevant for validating multilingual lyrics-matching aspects. 

\subsection{Baselines}
We consider an extensive set of baselines covering different approaches to lyrics-based matching. TF–IDF baselines follow Correya et al.~\cite{correya2018large}, comparing transcriptions either directly with cosine similarity or with a Lucene-style “More Like This” query (we denote these baselines as TF-IDF-Cosine and TF-IDF-Lucene, respectively). Another baseline encodes Whisper ASR transcriptions using Sentence-BERT (SBERT)~\cite{sbert} and compares them using both cosine similarity and a learnt Transformer using NT-Xent (we denote the two baselines as ASR-SBERT-Cosine and ASR-SBERT-Trasf, respectively). We also  include a representation-only baseline where we directly average Whisper decoder embeddings across time and compute similarities with cosine distance, without any further training (denoted as Whisper-AvgEmb). This baseline isolates the quality of raw representations extracted from Whisper before our proposed adaptation. Finally, we consider a random baseline that assigns distances uniformly at random, and a Non-instrumental Oracle that aims to account for instrumental versions that are not detectable through a lyrics-based approach. This oracle provides an upper bound by setting distances to zero only for same-clique songs with `valid' transcriptions, where validity is enforced through a dedicated process that cleans text and filters out unreliable cases associated to instrumental versions. For a song to be considered a match, the oracle requires it to have (i) a minimum length of 10 words, (ii) at least five alphanumeric characters, (iii) limited bigram/trigram repetition (no more than 70$\%$ repetition with at least three unique bigrams and two unique trigrams), (iv) no single phrase repeated in more than half of the text, and (v) no purely musical content such as annotations, humming, or repetitive syllable patterns.

\subsection{Training and Evaluation}
\label{sec:training-evaluation}
We train all models using the AdamW optimizer with a learning rate of 
$10^{-4}$, a weight decay of $10^{-3}$, a cosine schedule with 50 warmup epochs and a minimum learning rate 
of $10^{-6}$. Training is carried out for a maximum of 1000 
epochs with a batch size of 64 across four GPUs. To prevent overfitting, 
we apply early stopping with a patience of 20~epochs, monitoring mean 
average precision (MAP) on the validation set as the stopping criterion. We also use MAP to report results on the test set, which conforms to standard practice~\cite{yesiler_audio-based_2021}.

As mentioned, during training, we sample random subsequences of length $k=1500$ tokens from the 
Whisper representations $\textbf{H}$ to improve robustness and provide diverse 
temporal coverage of each track. However, at validation time, we adopt a 
deterministic procedure, always selecting the first $k$~tokens from each track to ensure consistency across epochs. 
At test time, we extract overlapping subsequences of $k$ tokens with 90\% 
overlap from both query and candidate tracks. 
Similarities for each query and candidate pair are computed on a full-track basis, using the two corresponding chunk-based embedding sequences and taking the maximum pairwise cosine similarity. 
This best-match approach simulates a 
chunk-based retrieval system that exploits the most discriminative 
portions of each track. 

        
        

\section{Results}
\label{sec:results}

Table~\ref{tab:results-WEALY} first highlights the performance of transcription-based baselines (TF-IDF and ASR-SBERT variants), which provide limited retrieval quality. The representation-only baseline (Whisper-AvgEmb) performs even worse, showing that averaged decoder embeddings do not generalise, and that a dedicated model is needed to learn effective representations. Overall, WEALY consistently outperforms all transcription-based methods, establishing a reproducible lyrics-aware baseline that scales with dataset size. 
For completeness, we here also note the results reported by~\cite{DBLP:conf/ismir/DuLZLWLZ24}, which introduces two Whisper-based variants: ``Whisper-AR'', described as an autoregressive feature extraction approach, and ``Whisper-E'', obtained by fine-tuning with adapters. The terminology in~\cite{DBLP:conf/ismir/DuLZLWLZ24} is somewhat ambiguous, but the reported MAP on SHS is 0.708 for Whisper-AR and 0.437 for Whisper-E. While a direct comparison is difficult due to the unclear methodology, our results are broadly aligned with these numbers, with the difference that WEALY offers a transparent and fully reproducible end-to-end pipeline.


Table~\ref{tab:ablation-wealy} reports the ablation study using SHS. First, the choice of loss function is critical: NT-Xent clearly outperforms both triplet loss and CLEWS loss, the latter being originally designed for multi-sequence similarity but proving less effective in this setting. Regarding pooling and architecture, simple averaging and the CLS token are competitive, but consistently below the GeM pooling baseline. In contrast, directly averaging Whisper embeddings and passing them through an MLP performs substantially worse, confirming that temporal modeling is essential for capturing lyric semantics. The language-constrained experiment further shows that multilinguality is a strength of Whisper: restricting decoding to English leads to a noticeable performance drop, indicating that multilingual cues in the latents provide useful information for retrieval. 

Finally, Table~\ref{tab:results-WEALY-multi} focuses more on MVI and demonstrates that multimodality brings clear improvements. A simple late-fusion of WEALY and CLEWS at the distance level yields a performance that surpasses both unimodal approaches, indicating that the two models capture complementary information. This improvement is obtained without introducing additional model complexity, relying only on a straightforward distance-level combination. While multimodal fusion and MVI is not the primary focus of this work, these results highlight the potential of lyrics matching for advancing the latter.

\begin{table}[t]
\vspace{-2mm}
\caption{Main results: MAP on the three considered datasets. }
\vspace{2mm}
\label{tab:results-WEALY}
\centering
\resizebox{\columnwidth}{!}{ 
\begin{tabular}{l|ccc}
\toprule
\textbf{Method} & \textbf{DVI} & \textbf{SHS} & \textbf{LYC} \\
\midrule
Random & 0.001 $\pm$ 0.000 & 0.003 $\pm$ 0.003 & 0.002 $\pm$ 0.002 \\
Non-instrumental Oracle & 0.967 $\pm$ 0.000 & 0.956 $\pm$ 0.004 & 0.954 $\pm$ 0.004\\
TF–IDF-Cosine & 0.272 $\pm$ 0.002 & 0.503 $\pm$ 0.008 & 0.537 $\pm$ 0.009 \\
TF–IDF-Lucene & 0.242 $\pm$ 0.002 & 0.457 $\pm$ 0.008 & 0.486 $\pm$  0.009 \\
ASR-SBERT-Cosine & 0.294 $\pm$ 0.002  & 0.508 $\pm$ 0.008 & 0.573 $\pm$  0.009 \\
ASR-SBERT-Trasf & N/A & 0.480 $\pm$ 0.001 &  0.516 $\pm$  0.008 \\
Whisper-AvgEmb & 0.166 $\pm$ 0.001 & 0.297 $\pm$ 0.007 & 0.322 $\pm$ 0.007\\
\textbf{WEALY} &  \textbf{0.328 $\pm$  0.002} & \textbf{0.640 $\pm$ 0.008} & \textbf{0.692 $\pm$ 0.008} \\
\bottomrule
\end{tabular}}
\vspace{-0.3cm} 
\end{table}


\begin{table}[t]
\caption{Ablation study of WEALY on the SHS dataset. 
}
\vspace{2mm}
\label{tab:ablation-wealy}
\centering
\resizebox{0.95\columnwidth}{!}{
\begin{tabular}{ll|c}
\toprule
\textbf{Category} & \textbf{Method} & \textbf{MAP} \\
\midrule
\textbf{Default} & \textbf{WEALY} & \textbf{0.640 $\pm$ 0.008} \\
\cdashline{1-3}\rule{-2pt}{2.5ex}
\multirow{2}{*}{Loss functions} 
   & WEALY - Triplet loss & 0.548 $\pm$ 0.008 \\
   & WEALY - CLEWS loss & 0.450 $\pm$ 0.008 \\
\cdashline{1-3}\rule{-2pt}{2.5ex}
\multirow{3}{*}{Pooling strategies} 
   & WEALY - Simple average & 0.627 $\pm$ 0.008 \\
   & WEALY - CLS token & 0.621 $\pm$ 0.008 \\
   & WEALY - Average+MLP & 0.389 $\pm$ 0.008 \\
\cdashline{1-3}\rule{-2pt}{2.5ex}
\multirow{1}{*}{Language settings} 
   & WEALY - English only & 0.578 $\pm$ 0.008 \\
\bottomrule
\end{tabular}}
\vspace{-0.3cm}
\end{table}

\begin{table}[!t]
\caption{Comparison of audio content-based MVI approaches with the proposed multimodal approach on the SHS dataset.}
\vspace{2mm}
\label{tab:results-WEALY-multi}
\centering
\resizebox{0.75\columnwidth}{!}{
\begin{tabular}{l|c}
\toprule
\textbf{Method} & \textbf{MAP} \\
\midrule
ByteCover1/2 (as reported by~\cite{serra2025supervised}) & 0.813 $\pm$ 0.006  \\ 
ByteCover3.5~\cite{DBLP:conf/ismir/DuLZLWLZ24} & 0.857\qquad\quad \,~~ \\
CLEWS~\cite{serra2025supervised} & 0.876 $\pm$ 0.005 \\
\textbf{WEALY+CLEWS} &  \textbf{0.912 $\pm$ 0.004}\\
\bottomrule
\end{tabular}}
\vspace{-0.3cm}
\end{table}



\section{Conclusion}
In this work, we introduced WEALY, the first reproducible end-to-end pipeline for audio-based lyrics matching. WEALY builds on Whisper decoder embeddings, which are extracted directly from raw audio without relying on intermediate transcriptions, and leverages the graound truth of musical version identification (MVI) datasets to train a transformer encoder via contrastive learning. This design allows us to capture lyrics-aware representations that are robust, scalable, and directly optimized for retrieval. Our experiments showed that WEALY consistently outperforms transcription-based pipelines, scales effectively to large datasets, and benefits from the latent multilingual properties of Whisper, which preserves useful cross-lingual cues for matching. While results remain below the strongest audio content-based models for MVI, we observed that fusing WEALY with them through a simple distance-level combination can lead to clear improvements, underlining the complementarity of lyric and audio cues, and pointing to promising multimodal extensions for MVI. Overall, our findings establish WEALY as a strong and transparent baseline for lyrics matching, with direct implications for music information retrieval tasks such as version identification, copyright detection, and music discovery. 

\section{Acknowledgments}
We acknowledge ISCRA for awarding this project access to the LEONARDO supercomputer, owned by the EuroHPC Joint Undertaking, hosted by CINECA (Italy).

\bibliographystyle{IEEEtran}
\bibliography{refs}

\begin{thebibliography}{10}
\providecommand{\url}[1]{#1}
\csname url@samestyle\endcsname
\providecommand{\newblock}{\relax}
\providecommand{\bibinfo}[2]{#2}
\providecommand{\BIBentrySTDinterwordspacing}{\spaceskip=0pt\relax}
\providecommand{\BIBentryALTinterwordstretchfactor}{4}
\providecommand{\BIBentryALTinterwordspacing}{\spaceskip=\fontdimen2\font plus
\BIBentryALTinterwordstretchfactor\fontdimen3\font minus \fontdimen4\font\relax}
\providecommand{\BIBforeignlanguage}[2]{{%
\expandafter\ifx\csname l@#1\endcsname\relax
\typeout{** WARNING: IEEEtran.bst: No hyphenation pattern has been}%
\typeout{** loaded for the language `#1'. Using the pattern for}%
\typeout{** the default language instead.}%
\else
\language=\csname l@#1\endcsname
\fi
#2}}
\providecommand{\BIBdecl}{\relax}
\BIBdecl

\bibitem{whisper}
A.~Radford, J.~W. Kim, T.~Xu, G.~Brockman, C.~McLeavey, and I.~Sutskever, ``Robust speech recognition via large-scale weak supervision,'' in \emph{International Conference on Machine Learning (ICML)}, 2023, pp. 28\,492--28\,518.

\bibitem{Balluff2024LyricCovers}
M.~Balluff, M.~Auch, P.~Mandl, and C.~Wolff, ``Lyriccovers 2.0: An enhanced dataset for cover song analysis,'' \emph{IADIS International Journal on WWW/Internet}, vol.~22, no.~2, pp. 75--92, 2024.

\bibitem{yesiler_audio-based_2021}
F.~Yesiler, G.~Doras, R.~M. Bittner, C.~J. Tralie, and J.~Serrà, ``Audio-based musical version identification: elements and challenges,'' \emph{IEEE Signal Processing Magazine}, vol.~38, no.~6, pp. 115--136, 2021.

\bibitem{knees2005multiple}
P.~Knees, M.~Schedl, and G.~Widmer, ``Multiple lyrics alignment: Automatic retrieval of song lyrics.'' in \emph{International Society for Music Information Retrieval Conference (ISMIR)}, 2005, pp. 564--569.

\bibitem{patra2017retrieving}
B.~G. Patra, D.~Das, and S.~Bandyopadhyay, ``Retrieving similar lyrics for music recommendation system,'' in \emph{Proceedings of the International Conference on Natural Language Processing (ICON)}, 2017, pp. 290--297.

\bibitem{serra2025supervised}
J.~Serr{\`a}, R.~O. Araz, D.~Bogdanov, and Y.~Mitsufuji, ``Supervised contrastive learning from weakly-labeled audio segments for musical version matching,'' in \emph{International Conference on Machine Learning (ICML)}, 2025.

\bibitem{bytecover2}
X.~Du, K.~Chen, Z.~Wang, B.~Zhu, and Z.~Ma, ``{ByteCover2}: Towards dimensionality reduction of latent embedding for efficient cover song identification,'' in \emph{IEEE International Conference on Acoustics, Speech and Signal Processing (ICASSP)}, 2022, pp. 616--620.

\bibitem{bytecover3}
X.~Du, Z.~Wang, X.~Liang, H.~Liang, B.~Zhu, and Z.~Ma, ``{ByteCover3}: Accurate cover song identification on short queries,'' in \emph{IEEE International Conference on Acoustics, Speech and Signal Processing (ICASSP)}, 2023, pp. 1--5.

\bibitem{vaglio-hal-03356164}
A.~Vaglio, R.~Hennequin, M.~Moussallam, and G.~Richard, ``{The Words Remain The Same: Cover Detection with Lyrics Transcription},'' in \emph{{International Society for Music Information Retrieval Conference (ISMIR)}}, 2021.

\bibitem{correya2018large}
A.~Correya, R.~Hennequin, and M.~Arcos, ``Large-scale cover song detection in digital music libraries using metadata, lyrics and audio features,'' \emph{arXiv preprint arXiv:1808.10351}, 2018.

\bibitem{DBLP:conf/ismir/DuLZLWLZ24}
X.~Du, M.~Liu, P.~Zou, X.~Liang, Z.~Wang, H.~Liang, and B.~Zhu, ``X-cover: Better music version identification system by integrating pretrained {ASR} model,'' in \emph{International Society for Music Information Retrieval Conference (ISMIR)}, 2024, pp. 70--77.

\bibitem{DBLP:conf/ismir/ZhuoYPMLZLDFLBC23}
L.~Zhuo, R.~Yuan, J.~Pan, Y.~Ma, Y.~Li, G.~Zhang, S.~Liu, R.~B. Dannenberg, J.~Fu, C.~Lin, E.~Benetos, W.~Chen, W.~Xue, and Y.~Guo, ``Lyricwhiz: Robust multilingual zero-shot lyrics transcription by whispering to chatgpt,'' in \emph{International Society for Music Information Retrieval Conference (ISMIR)}, 2023, pp. 343--351.

\bibitem{radenovic2018fine}
F.~Radenovi{\'c}, G.~Tolias, and O.~Chum, ``Fine-tuning {CNN} image retrieval with no human annotation,'' \emph{IEEE Transactions on Pattern Analysis and Machine Intelligence}, vol.~41, no.~7, pp. 1655--1668, 2018.

\bibitem{pmlr-v119-chen20j}
T.~Chen, S.~Kornblith, M.~Norouzi, and G.~Hinton, ``A simple framework for contrastive learning of visual representations,'' in \emph{International Conference on Machine Learning (ICML)}, 2020, pp. 1597--1607.

\bibitem{Schroff_2015_CVPR}
F.~Schroff, D.~Kalenichenko, and J.~Philbin, ``Facenet: A unified embedding for face recognition and clustering,'' in \emph{Proceedings of the IEEE Conference on Computer Vision and Pattern Recognition (CVPR)}, 2015.

\bibitem{araz2024discogs}
R.~O. Araz, X.~Serra, and D.~Bogdanov, ``{Discogs-VI}: A musical version identification dataset based on public editorial metadata,'' in \emph{International Society for Music Information Retrieval Conference (ISMIR)}, 2024.

\bibitem{yu2020learning}
Z.~Yu, X.~Xu, X.~Chen, and D.~Yang, ``Learning a representation for cover song identification using convolutional neural network,'' in \emph{IEEE International Conference on Acoustics, Speech and Signal Processing (ICASSP)}, 2020, pp. 541--545.

\bibitem{sbert}
N.~Reimers and I.~Gurevych, ``Sentence-{BERT}: Sentence embeddings using {S}iamese {BERT}-networks,'' in \emph{Conference on Empirical Methods in Natural Language Processing and International Joint Conference on Natural Language Processing (EMNLP-IJCNLP)}, 2019, pp. 3982--3992.

\end{thebibliography}

\end{document}